\documentclass[letterpaper]{article} % DO NOT CHANGE THIS
\usepackage[]{aaai24}  % DO NOT CHANGE THIS
\usepackage{aaai24}  % DO NOT CHANGE THIS
\usepackage{times}  % DO NOT CHANGE THIS
\usepackage{helvet}  % DO NOT CHANGE THIS
\usepackage{courier}  % DO NOT CHANGE THIS
\usepackage[hyphens]{url}  % DO NOT CHANGE THIS
\usepackage{graphicx} % DO NOT CHANGE THIS
\usepackage{natbib}  % DO NOT CHANGE THIS AND DO NOT ADD ANY OPTIONS TO IT
\usepackage{caption} % DO NOT CHANGE THIS AND DO NOT ADD ANY OPTIONS TO IT
\usepackage{multirow}
\usepackage{algorithm}
\usepackage{algorithmic}

\usepackage{newfloat}
\usepackage{listings}
\DeclareCaptionStyle{ruled}{labelfont=normalfont,labelsep=colon,strut=off} % DO NOT CHANGE THIS
\floatstyle{ruled}
\newfloat{listing}{tb}{lst}{}
\floatname{listing}{Listing}
\title{Brain-optimized inference improves reconstructions of fMRI brain activity}

\author{
    Reese Kneeland\textsuperscript{\rm 1},
    Jordyn Ojeda\textsuperscript{\rm 1},
    Ghislain St-Yves\textsuperscript{\rm 2},
    Thomas Naselaris\textsuperscript{\rm 2}
    }

\affiliations {
    \textsuperscript{\rm 1}Department of Computer Science, University of Minnesota\\
    \textsuperscript{\rm 2}Department of Neuroscience, University of Minnesota \\
    \{rek, ojeda040, gstyves, nase0005\}@umn.edu
}
\begin{document}{

\maketitle

\begin{abstract}
The release of large datasets and developments in AI have led to dramatic improvements in decoding methods that reconstruct seen images from human brain activity. We evaluate the prospect of further improving recent decoding methods by optimizing for consistency between reconstructions and brain activity during inference. We sample seed reconstructions from a base decoding method, then iteratively refine these reconstructions using a brain-optimized encoding model that maps images to brain activity.  At each iteration, we sample a small library of images from an image distribution (a diffusion model) conditioned on a seed reconstruction from the previous iteration. We select those that best approximate the measured brain activity when passed through our encoding model, and use these images for structural guidance during the generation of the small library in the next iteration. We reduce the stochasticity of the image distribution at each iteration, and stop when a criterion on the “width” of the image distribution is met. We show that when this process is applied to recent decoding methods, it outperforms the base decoding method as measured by human raters, a variety of image feature metrics, and alignment to brain activity. These results demonstrate that reconstruction quality can be significantly improved by explicitly aligning decoding distributions to brain activity distributions, even when the seed reconstruction is output from a state-of-the-art decoding algorithm. Interestingly, the rate of refinement varies systematically across visual cortex, with earlier visual areas generally converging more slowly and preferring narrower image distributions, relative to higher-level brain areas. Brain-optimized inference thus offers a succinct and novel method for improving reconstructions and exploring the diversity of representations across visual brain areas.

\end{abstract}

\section{Introduction}
Reconstructing visual stimuli from measured brain activity is a long-standing problem in computational neuroscience. Improvements bring us closer to understanding visual representation in the brain by providing interpretable tools for examining brain states \cite{naselaris2011, St-Yves_heirarchy}. 

Two recent developments have accelerated progress in image reconstruction from human brain activity: large datasets that offer samples of brain activity in response to many thousands of natural scenes \cite{Allen2021a}, and the open-sourcing of powerful image-generators that can invert abstract feature representations into distributions over natural scenes \cite{stablediffusion}. In many recent approaches, this data is used to train a decoding model that maps brain activity directly to feature representations that are fed into the generator. The decoding model is typically trained by minimizing the distance between its outputs and the ground-truth feature representation.

Although this approach has been proven quite successful \cite{St-Yves_gan, lin2022mind, gu2023decoding, Takagi2022.11.18.517004, takagi2023improving, ozcelik2023braindiffuser, lu2023minddiffuser, scotti2023reconstructing}, it ignores an important constraint that could be leveraged to improve reconstructions: If a reconstruction is faithful to the image encoded in brain activity, then it should accurately predict brain activity when input to an accurate encoding model \cite{cycleconsistency}. Because encoding models provide a more comprehensive analysis of the proximity between images and brain activity \cite{naselaris2011}, an ideal decoding model would not only minimize the expected error of a decoded feature representation for images in the training set, but also maximize alignment with the expected pattern of brain activity for each image in the test set. Because this alignment can be measured for each test item, it could potentially be used to assess reconstruction quality when the ground-truth image is unknown (e.g., when reconstructing a mental image). Given that human judgment is grounded in human brain activity, it could be the case that optimizing for alignment with brain activity could improve reconstructions relative to the judgments of human observers, whose assessments are often in disagreement with feature similarity metrics. 

In addition to aligning reconstructions with expected patterns of brain activity, an ideal decoding method would also output a distribution that was aligned to variance in brain activity across repetitions of the same image. Such alignment could, in principle, help to balance semantic and structural determination, protect against overfitting to noise in the brain data or errors in the encoding model, and potentially capture a representation of perceptual uncertainty \cite{sensoryuncertainty}. Decoding methods that generate reconstructions using a diffusion model offer a convenient handle on the stochasticity of the image distribution in the form of a "strength" parameter, which controls the relative influence of high-level semantic and low-level structural information on the reconstruction. When the strength parameter is set to high values, the image distribution is more stochastic, as it implicitly assigns more uniform probabilities to any image consistent with it's semantic guidance. By manipulating the strength parameter, we can control the amount of variability in the outputs of the encoding model through which we pass the reconstructions.

We describe a stochastic search algorithm that independently optimizes reconstructions of each image in the test set, using the correlation between measured brain activity and the outputs of an encoding model as a target. We iteratively refine a \textit{distribution} of images, and implement a  stopping criterion for the distribution width that utilizes the neural variance in a set of brain activity patterns. This algorithm runs during the inference stage on top of a direct decoding method. We refer to this process as brain-optimized inference (BOI).

% this subsection is now redundant. it was also poorly written.
% \subsection{Brain correlation as an optimization objective}
% \label{braincorrelation}

%  The use of encoding models to improve reconstruction quality is currently a missed opportunity for enhancing reconstruction quality. Encoding models can be used to examine a more comprehensive relationship between images and the brain \cite{naselaris2011}, and by using them to make our reconstructions more compatible with the brain, we might also make them more compatible with human subjective assessments. To take advantage of this opportunity, we utilize a "brain correlation" metric, which uses a voxel-wise brain encoding model \cite{St-Yves_heirarchy} to map images into the space of brain activity, allowing for the scoring of reconstructions based on their Pearson correlation with the original pattern of brain activity. By treating the space of brain activity as a latent space in which we can measure accuracy statistics, this method circumvents the need for a ground truth image to score against, instead using a target that should in principle be more faithful to the perceptual representation of the observer, which can often be a lossy and incomplete representation of the external sensory input. In principle, this metric could also allow for the optimization and evaluation of reconstructions in cases where a ground truth stimulus is not available, such as in mental imagery or dreams. 

\subsection{Overview of brain-optimized inference procedure}
\label{boi}
Our approach starts from the decoded latent variables provided by a direct decoding method. For all results in this paper, we utilize the MindEye reconstruction method \cite{scotti2023reconstructing} as the base direct decoding method, although in principle our algorithm is compatible with any direct decoding approach that uses a diffusion model. After generating a small library of images using the decoded latent variables, we utilize a brain-optimized encoding model to perform an optimization step that minimizes loss in the space of brain activity using a brain correlation metric, which passes images through a voxel-wise encoding model to obtain predicted patterns of brain activity $\beta'$, which are subsequently compared to the averaged trial repetitions of brain activity evoked by the target image $\beta$ via a Pearson correlation. During the optimization process, we use this metric to select images from the library that are closely aligned to the brain activity pattern evoked by the target image. By using encoding models optimized to predict activity in most of the visual cortex, we leverage the rich diversity of representations in the human visual system to pick images that are holistically aligned to both structural and semantic features. The selected images are then used as structural guidance in the generation of a new library of images, addressing the structural determinism limitation by providing our structural guidance iteratively over time. The width of the generated image distribution is progressively reduced across iterations until it meets a stopping criterion for the distribution width, representing our admitted stochasticity directly in the output. To obtain our stopping criterion and avoid overfitting, we first train a denoising autoencoder, designed to map normalized fMRI voxel responses $\beta$ to the noise-free predicted voxel responses $\beta'$ generated by passing the ground truth images through our encoding model. This produces denoised patterns of brain activity $\hat{\beta}$. Our stopping criterion is then defined as the point when the across-trial variance of the denoised brain activity matches the variance of the output distribution of predicted brain activity patterns $\beta'$ generated from passing the output image distribution through an encoding model. This condition is represented in the following equation:
$$\sigma(\hat{\beta}) = \sigma(\beta')$$

This stopping criterion is a preliminary attempt to capture some of the sample-specific brain variance in the fMRI data while avoiding overfitting to our encoding model. Future work could explore stopping criteria that better capture the perceptual uncertainty of the observer, as our current approach destroys a large portion of the brain noise present in the original betas in the denoising process in service of matching the noise-free characteristics of our encoding model.

\subsection{Overview of the results}

We show that upon reaching the stopping criterion for the width of the distribution, the resulting brain-aligned image distribution yields an improvement in performance as assessed by human raters, most traditional image feature metrics, and brain correlation scores when compared to the base decoding method. Full results can be seen in Section \ref{results}.

Furthermore, by analyzing search dynamics in Section \ref{analysis} we observe that a different subset of samples from our output distributions meets our target variance threshold for different brain areas and that they converge at different rates. We show how this degeneracy increases across visual cortex, as evidenced by the rapid rate at which reconstructions align to activity in high-level areas, relative to low-level areas. We argue that “rate of convergence” is a succinct and attractive metric of representational invariance across brain areas and brain states.

\section{Methods}

\subsection{Dataset}
We use the Natural Scenes Dataset (NSD) \cite{Allen2021a}, which comprises $27,750$ scans distributed across the $37$ sessions publicly released by NSD for subjects $1, 2, 5,$ and $7$, which are the subjects that completed all scanning sessions. We partition the data for each subject into three distinct sets: training $(n=21,216)$, validation $(n=3,764)$, and test $(n=2,770)$. The test set is derived from a shared pool of $982$ images that all subjects have seen, whereas the training and validation sets are generated from the remaining data. All images in the dataset are from the Microsoft COCO dataset \cite{microsoftcoco}. In this investigation, we mask preprocessed fMRI signals using the nsdgeneral mask in 1.8mm resolution. The ROI consists of $[15724,14278,13039,12682]$ voxels for the four subjects and includes the entire visual cortex region. For further information regarding data preprocessing, please refer to \cite{Allen2021a}. For some of our analysis, we further segment the nsdgeneral ROI into V1, V2, V3, V4 (collectively called early visual cortex) and higher visual areas (the set complement of nsdgeneral and early visual cortex).

\subsection{Models}
This method primarily utilizes pre-trained implementations of open-source models, the most vital of which is the GNet brain-optimized voxel-wise encoding model \cite{St-Yves_heirarchy}. As a brain encoding model with sufficient sensitivity to both structural and semantic image features, we use it to generate predictions of brain activity $\beta'$ from candidate images during the inference process. To train the denoising autoencoder for our stopping criterion, the encoded $\beta'$s corresponding to the ground truth images serve as targets in conjunction with the original fMRI $\beta$s to produce a denoised $\hat{\beta}$ with the same noise profile as the encoded $\beta'$s. For image generation during the brain-optimized inference algorithm, this implementation utilizes the same pre-trained Versatile Diffusion model \cite{Xu_2023_ICCV} as MindEye \cite{scotti2023reconstructing}, although in principle our method is compatible with any diffusion model that can accept semantic and structural guidance, in addition to a strength parameter to modulate them.

\begin{figure}[!htb]
\begin{center}
\includegraphics[width=\columnwidth]{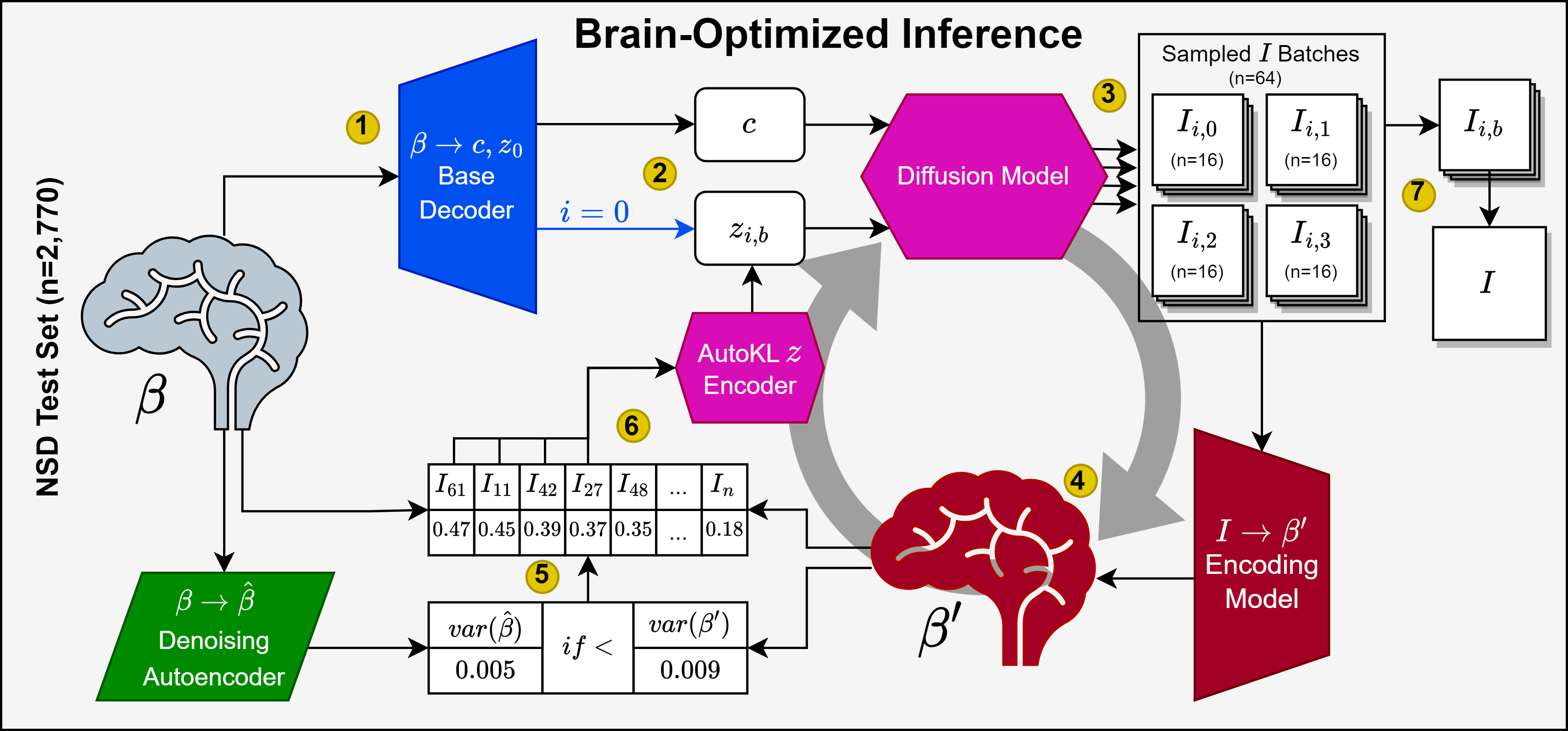}
\end{center}
\caption{Pipeline diagram for our method. The initial outputs from the MindEye decoder (blue box) produce the CLIP vector $c$ and an initial latent $z$ vector to seed the rest of the pipeline, in which we iteratively align an image distribution to the measured brain activity. The numbered components of the pipeline are detailed in Section \ref{search}.} 
\label{figure:pipeline}
\end{figure}

\subsection{Brain-optimized inference algorithm}
\label{search}
The stochastic search algorithm for our brain-optimized inference stage is designed to progressively reduce the width of a provided image distribution described by diffusion parameters while optimizing the search toward alignment with the measured brain activity. Versatile Diffusion exposes a strength parameter $\lambda$ that modulates the quantity of noise introduced to the latent $z$ vector, with a value of 1 representing the addition of 100\% noise and 0 signifying the addition of 0\% noise. Furthermore, the strength parameter dictates the number of CLIP-guided denoising steps to be executed, as $z$ vectors with reduced noise necessitate less denoising. This parameter effectively controls the breadth of the distribution of generated images originating from a given $z$ vector, a mechanism we utilize to control the distribution width \cite{kneeland2023reconstructing}. The following numbered items refer to Figure \ref{figure:pipeline}.

\begin{enumerate}
\item Our pipeline begins by applying a base decoding process to obtain a CLIP vector $c$ \cite{radford2021learning}, representing a point estimate of the decoded semantic content, and a latent $z_0$ vector, representing a point estimate of the low-level structural content. For this implementation we use the MindEye reconstruction method \cite{scotti2023reconstructing} as our base decoder.
\item High-level ($c$) and low-level ($z_0$) representations are input to the image generator. Iterations $i > 0$ are seeded with $z_{i,b}$ instead of $z_0$.
\item Image batches ($I_{i, b}$) are sampled from Versatile Diffusion conditioned on $c$ and $z_{i, b}$. Iterations $i > 0$ yield multiple batches of images, each generated using a different latent variable $z_{i,b}$ generated from one of the top $n_b$ images in the previous iteration. $0 <= i < 10 $, $0 <= b < n_b$, $n_b=4$. 
\item  Images in each batch are passed through the encoding model to yield predicted brain activity patterns $\beta’$. These $\beta’$s are compared against the original $\beta$s via a Pearson correlation to yield a brain correlation score for each image.
\item The original $\beta$ repetitions are autoencoded into $\hat{\beta}$s that match the variance properties of the encoded $\beta'$s. If the variance of the current batch of $\beta'$s is still larger than the variance of the $\hat{\beta}$s, we continue the search for another iteration.
\item The four images with the highest brain correlation scores are selected as the batch seeds for the next iteration. By using the top four images to generate four batches, we ensure our search does not get stuck in a local maximum. The four images are embedded into their corresponding $z_{i,b}$ vectors, each seeding $n_b$ images in the forthcoming batch $I_{i,b}$. 
\item When the stopping criterion is met in step $5$, the image distribution $I_{i,b}$ from the batch that yields the highest average brain correlation from that iteration is selected as the output distribution, from which the image with the highest brain correlation $I$ is selected as the output image displayed in Figure \ref{figure:papercomp}.
\end{enumerate}

\section{Results}
\label{results}
\subsection{Assessment of reconstruction quality}

\begin{figure}[!htb]
\begin{center}
\includegraphics[width=\columnwidth]{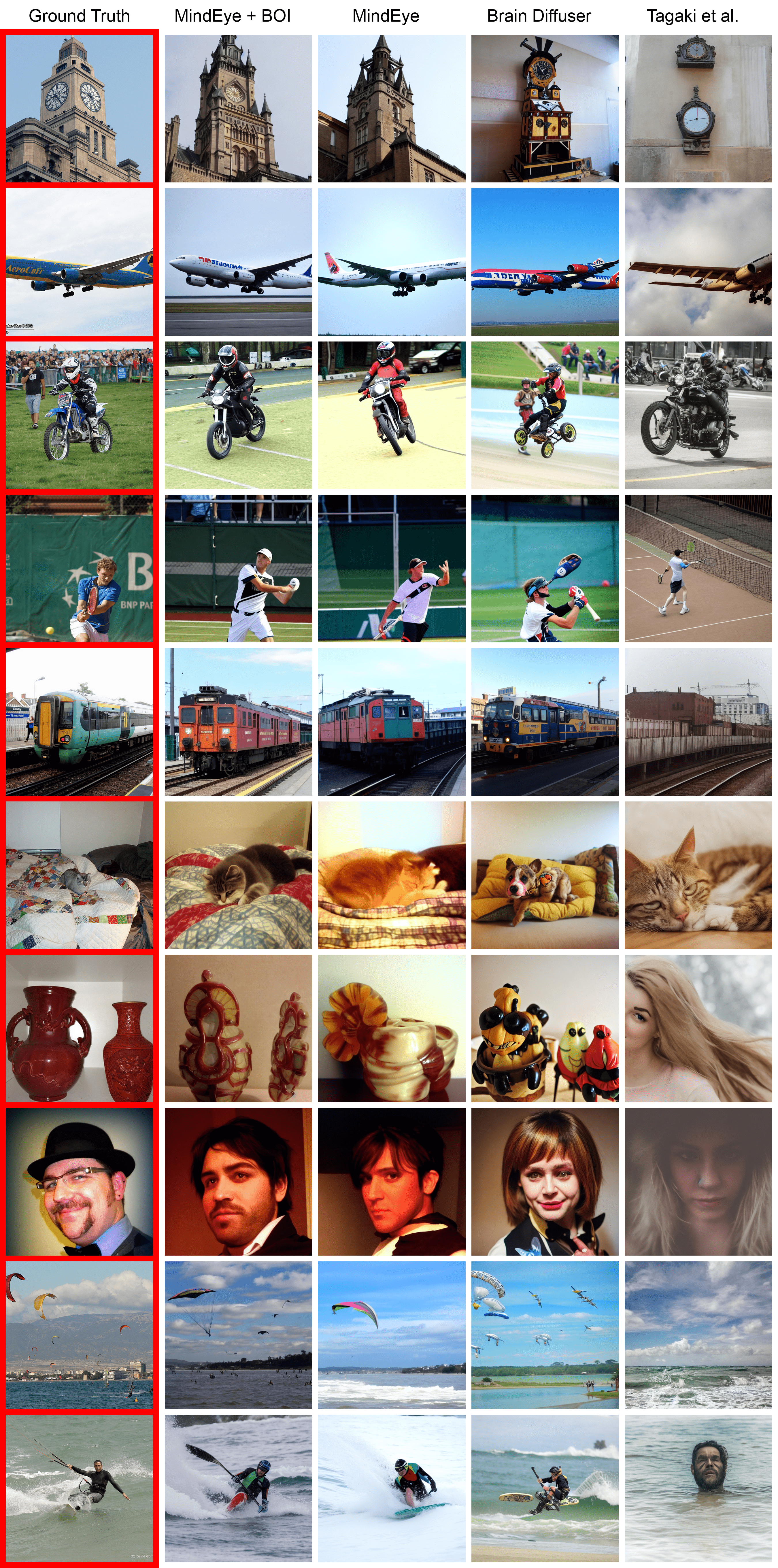}
\end{center}
\caption{Comparative assessment of reconstruction methods for subject 1. The first column is the ground truth image, indicated by the red border, while the second column is the reconstruction from our brain-optimized inference stage on top of the MindEye decoding method. The remaining rows represent results from previous decoding methods, including MindEye \cite{scotti2023reconstructing}, Brain Diffuser \cite{ozcelik2023braindiffuser}, and the "+Decoded Text" method from Tagaki et al. \cite{takagi2023improving}.}
\label{figure:papercomp}
\end{figure}

\begin{figure}[!htb]
\begin{center}
\includegraphics[width=\columnwidth]{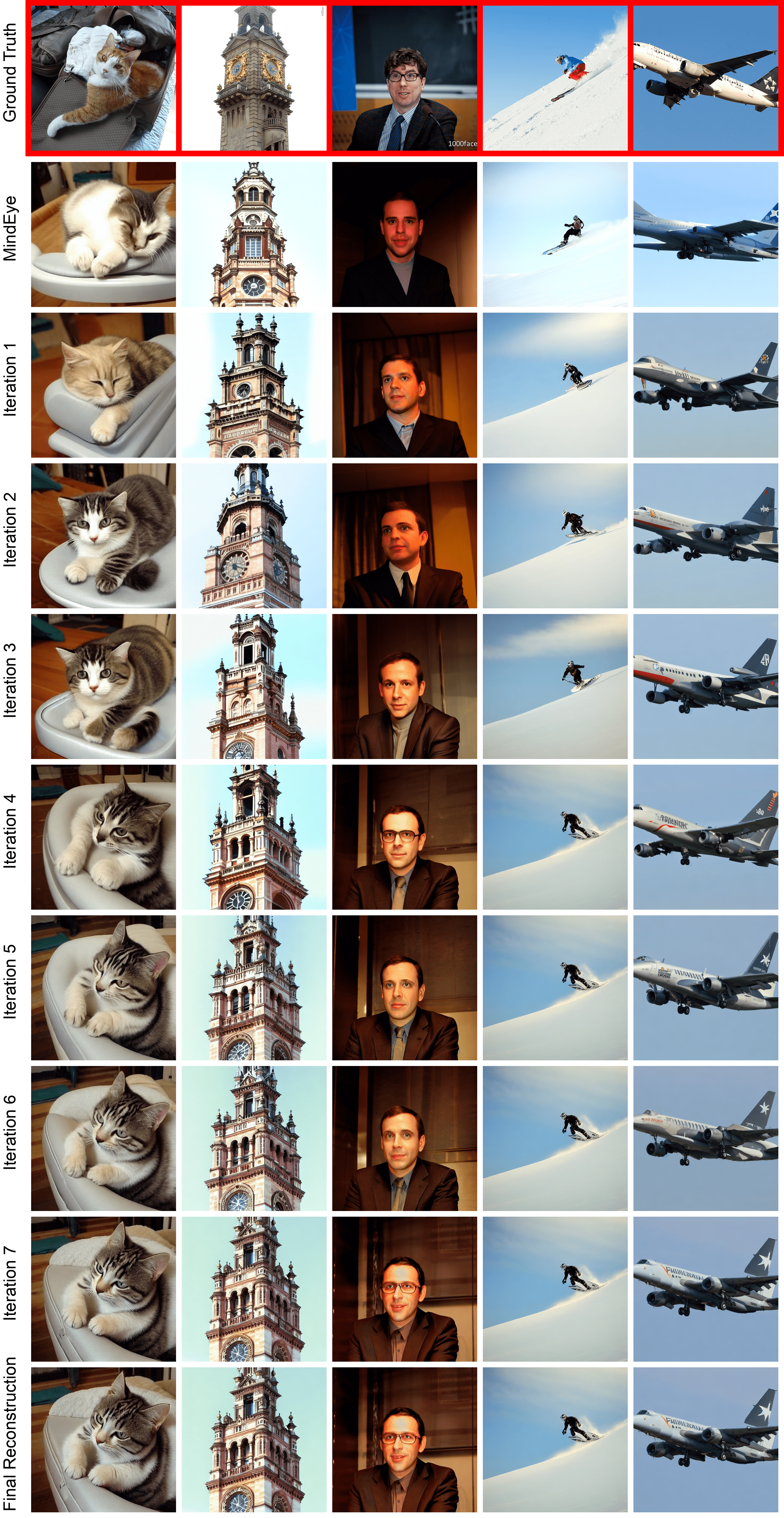}
\end{center}
\caption{Iterative analysis of high-quality reconstructions generated by our brain-optimized inference stage on top of the MindEye decoding method. The first row is the ground truth image, indicated by the red border, the second row is the reconstruction provided by the MindEye reconstruction method, and the remaining columns are the reconstructions produced at iterative stages of the BOI process. The last row is an image from the last output distribution of our MindEye + BOI method. These particular samples converged at or before iteration $7$.} 
\label{figure:iterationcomp}
\end{figure}

Figure \ref{figure:papercomp} presents a comparison of our "MindEye + BOI" reconstructions against several recent works, including MindEye \cite{scotti2023reconstructing}, Brain Diffuser \cite{ozcelik2023braindiffuser}, and the "+Decoded Text" method from Tagaki et al. \cite{takagi2023improving}. As shown, in the best case reconstructions recapitulate both the semantic content of images and their low-level structural details. At a glance, our reconstructions compare favorably to past methods. Figure \ref{figure:iterationcomp} shows intermediate images from the iterative stages of our optimization algorithm, illustrating how our algorithm provides iterative improvements to the images while slowly reducing the distribution width, as the images become more similar over time.

To validate the performance of our method, we conducted two subjective experiments (human raters, n=$54$), with our results shown in Figure \ref{figure:humanscores}. The first experiment was a two-way identification task, in which subjects were asked to select whether the image reconstructed from fMRI using our "MindEye + BOI" reconstruction method was more similar to the corresponding ground truth image than a randomly sampled reconstructed image from the test set (n=982). An example of the task can be seen in Figure \ref{figure:taskexample}. Subjects in this experiment identified the corresponding reconstruction as being more similar $95.62\%$ of the time ($p<0.001$). This establishes a new SOTA for human-rated image similarity, as the only other paper to perform such an experiment was the original method proposed by Takagi et al. \cite{Takagi2022.11.18.517004}, whose method achieved $84.29\%$. That method is notably different from the "+Decoded Text" method we compare against in Figure \ref{table:results}, which was released in a later technical report \cite{takagi2023improving}, and which does not have human subjective data. The second subjective experiment asked subjects to perform the same task, but with the "MindEye + BOI" reconstructions pitted directly against reconstructions from the base MindEye decoding method \cite{scotti2023reconstructing}. This experiment, designed to evaluate whether this additional brain-optimized inference stage improved the resulting reconstructions, demonstrated that the +BOI reconstructions were preferred $56.87\%$ of the time ($p<0.001$). Both of these experiments were conducted for all $982$ samples in the test set, for all $4$ subjects, for a total of $3,928$ trials each distributed randomly across the $54$ subjects. We believe subjective human evaluation to be the most comprehensive way to evaluate image similarity, a metric for which we establish a new SOTA.

\begin{figure}[!htb]
\begin{center}
\includegraphics[width=\columnwidth]{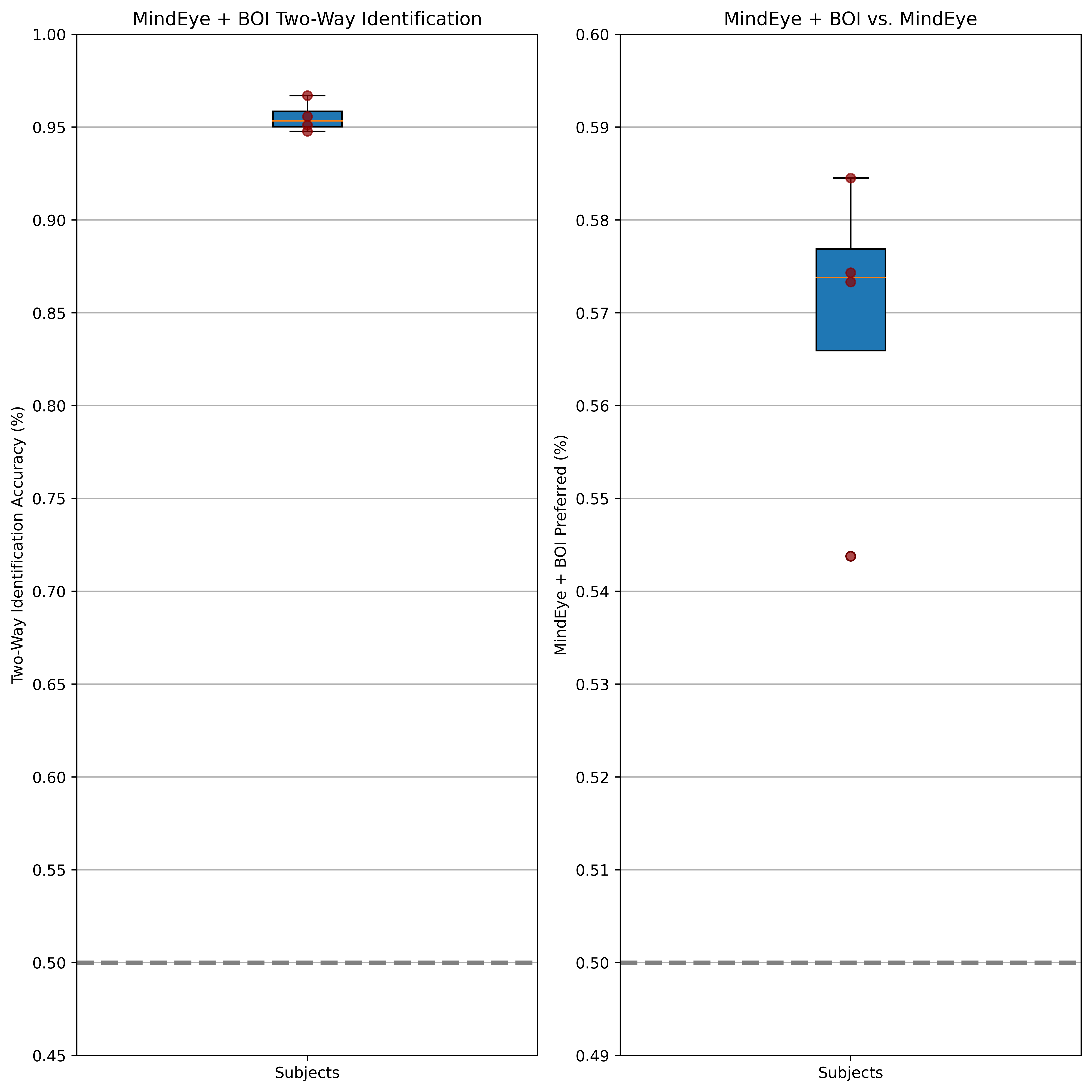}
\end{center}
\caption{Box and whisker plots for the subjective analysis experiments conducted with human raters (n=54). The dashed lines represent the chance threshold. The left plot is for a two-way identification task between our "MindEye + BOI" reconstruction and a random reconstruction. The right plot is for a head-to-head preference task between our "MindEye + BOI" reconstruction and the base MindEye reconstruction.}
\label{figure:humanscores}
\end{figure}

\begin{figure}[!htb]
\begin{center}
\includegraphics[width=\columnwidth]{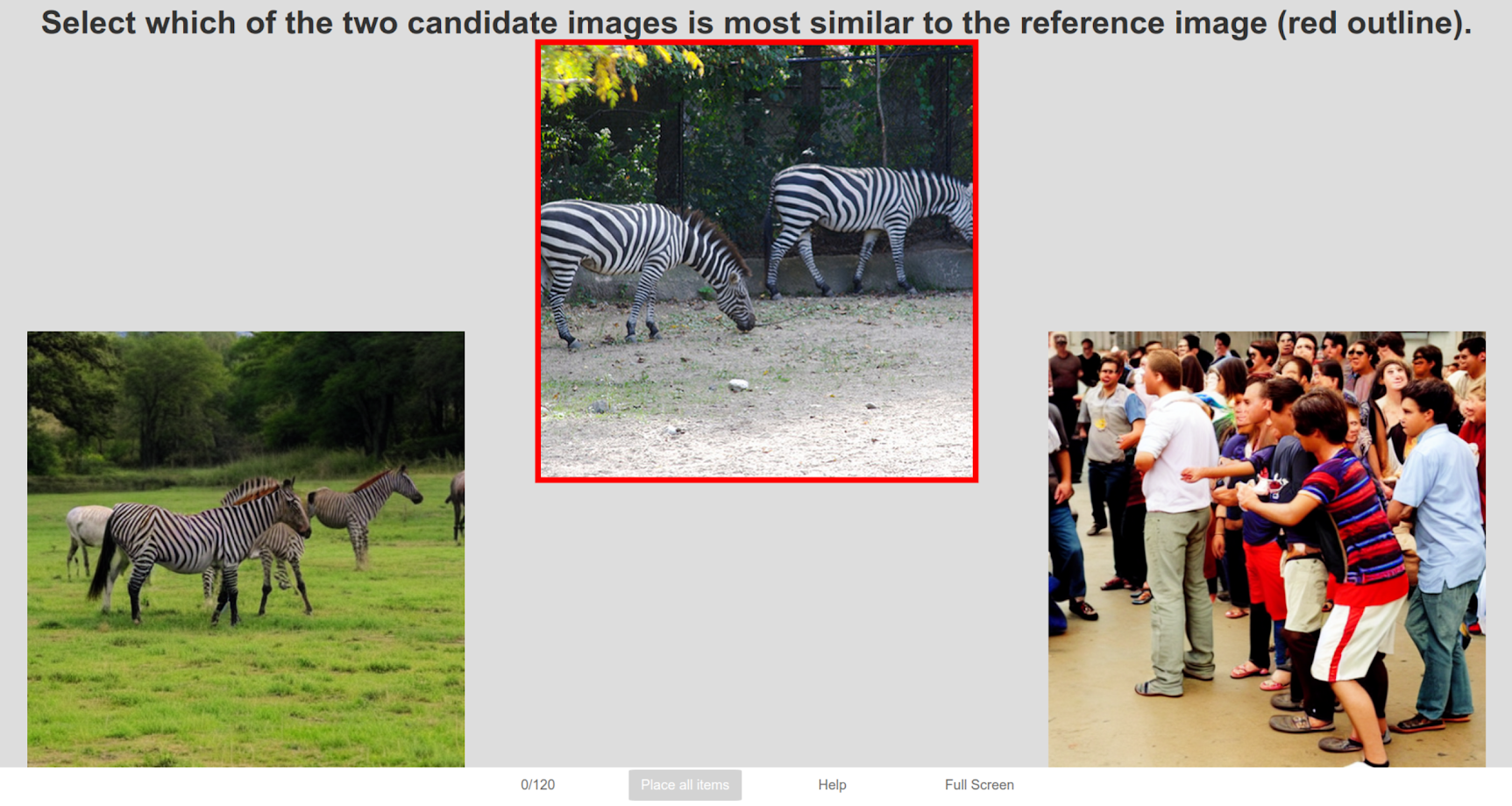}
\end{center}
\caption{An example of the task for the subjective experiments performed by human raters.}
\label{figure:taskexample}
\end{figure}

To further our analysis of reconstruction quality across different approaches, we calculated the proximity of our and others' reconstructions to ground truth in a variety of feature spaces and the space of brain activity (Table \ref{table:results}). To calculate the metrics for the other methods, we utilized the open-source implementations of each method to obtain 5 repetitions of the output images for the shared1000 test set. The numbers for MindEye + BOI (ours) are calculated on 5 images sampled from the output distribution of our brain-optimized inference algorithm. Although no additional optimization in feature space was performed, our reconstructions establish a new SOTA within many high-level feature spaces including the 7th layer of AlexNet, CLIP, InceptionV3, and EffNet-B. Our reconstructions also establish SOTA brain correlation scores for all visual ROIs we examined, an expected result given that our method explicitly optimizes for this metric during the inference process. 

% \usepackage{multirow}
% \usepackage{graphicx}
\begin{table}[!htb]
\centering
\resizebox{\columnwidth}{!}{%
\begin{tabular}{|clrrrr|}
\hline
\multicolumn{6}{|c|}{Quantitative Evaluation of Reconstruction Methods} \\ \hline
\multicolumn{1}{|c|}{Type} &
  \multicolumn{1}{c|}{Metric} &
  \multicolumn{1}{c|}{MindEye + BOI} &
  \multicolumn{1}{c|}{MindEye} &
  \multicolumn{1}{c|}{Brain Diffuser} &
  \multicolumn{1}{c|}{Tagaki et al.} \\ \hline
\multicolumn{1}{|c|}{Human} &
  \multicolumn{1}{l|}{2-way Identification} &
  \multicolumn{1}{r|}{\textbf{95.62\%}} &
  \multicolumn{1}{r|}{-} &
  \multicolumn{1}{r|}{-} &
  - \\ \hline
\multicolumn{1}{|c|}{\multirow{4}{*}{\rotatebox[origin=c]{90}{Low Level}}} &
  \multicolumn{1}{l|}{PixCorr} &
  \multicolumn{1}{r|}{0.259} &
  \multicolumn{1}{r|}{\textbf{0.310}} &
  \multicolumn{1}{r|}{0.266} &
  0.239 \\
\multicolumn{1}{|c|}{} &
  \multicolumn{1}{l|}{SSIM} &
  \multicolumn{1}{r|}{0.329} &
  \multicolumn{1}{r|}{0.331} &
  \multicolumn{1}{r|}{0.340} &
  \textbf{0.377} \\
\multicolumn{1}{|c|}{} &
  \multicolumn{1}{l|}{AlexNet(2)} &
  \multicolumn{1}{r|}{93.92\%} &
  \multicolumn{1}{r|}{\textbf{94.69\%}} &
  \multicolumn{1}{r|}{93.90\%} &
  78.47\% \\
\multicolumn{1}{|c|}{} &
  \multicolumn{1}{l|}{AlexNet(5)} &
  \multicolumn{1}{r|}{97.71\%} &
  \multicolumn{1}{r|}{\textbf{97.75\%}} &
  \multicolumn{1}{r|}{96.52\%} &
  85.52\% \\ \hline
\multicolumn{1}{|c|}{\multirow{5}{*}{\rotatebox[origin=c]{90}{High Level}}} &
  \multicolumn{1}{l|}{AlexNet(7)} &
  \multicolumn{1}{r|}{\textbf{97.53\%}} &
  \multicolumn{1}{r|}{97.44\%} &
  \multicolumn{1}{r|}{95.57\%} &
  87.26\% \\
\multicolumn{1}{|c|}{} &
  \multicolumn{1}{l|}{CLIP(2-way)↑} &
  \multicolumn{1}{r|}{\textbf{93.85\%}} &
  \multicolumn{1}{r|}{93.84\%} &
  \multicolumn{1}{r|}{91.70\%} &
  82.64\% \\
\multicolumn{1}{|c|}{} &
  \multicolumn{1}{l|}{Inception V3↑} &
  \multicolumn{1}{r|}{\textbf{93.88\%}} &
  \multicolumn{1}{r|}{93.81\%} &
  \multicolumn{1}{r|}{90.81\%} &
  83.24\% \\
\multicolumn{1}{|c|}{} &
  \multicolumn{1}{l|}{EffNet-B↓} &
  \multicolumn{1}{r|}{\textbf{0.645}} &
  \multicolumn{1}{r|}{0.647} &
  \multicolumn{1}{r|}{0.739} &
  0.817 \\
\multicolumn{1}{|c|}{} &
  \multicolumn{1}{l|}{SwAV↓} &
  \multicolumn{1}{r|}{0.367} &
  \multicolumn{1}{r|}{\textbf{0.366}} &
  \multicolumn{1}{r|}{0.430} &
  0.510 \\ \hline
\multicolumn{1}{|c|}{\multirow{6}{*}{\rotatebox[origin=c]{90}{Brain Corr.}}} &
  \multicolumn{1}{l|}{V1} &
  \multicolumn{1}{r|}{\textbf{0.386}} &
  \multicolumn{1}{r|}{0.333} &
  \multicolumn{1}{r|}{0.341} &
  0.166 \\
\multicolumn{1}{|c|}{} &
  \multicolumn{1}{l|}{V2} &
  \multicolumn{1}{r|}{\textbf{0.385}} &
  \multicolumn{1}{r|}{0.318} &
  \multicolumn{1}{r|}{0.323} &
  0.143 \\
\multicolumn{1}{|c|}{} &
  \multicolumn{1}{l|}{V3} &
  \multicolumn{1}{r|}{\textbf{0.389}} &
  \multicolumn{1}{r|}{0.309} &
  \multicolumn{1}{r|}{0.314} &
  0.142 \\
\multicolumn{1}{|c|}{} &
  \multicolumn{1}{l|}{V4} &
  \multicolumn{1}{r|}{\textbf{0.365}} &
  \multicolumn{1}{r|}{0.305} &
  \multicolumn{1}{r|}{0.305} &
  0.157 \\
\multicolumn{1}{|c|}{} &
  \multicolumn{1}{l|}{Higher Visual Cortex} &
  \multicolumn{1}{r|}{\textbf{0.406}} &
  \multicolumn{1}{r|}{0.353} &
  \multicolumn{1}{r|}{0.357} &
  0.251 \\
\multicolumn{1}{|c|}{} &
  \multicolumn{1}{l|}{Whole Visual Cortex} &
  \multicolumn{1}{r|}{\textbf{0.416}} &
  \multicolumn{1}{r|}{0.356} &
  \multicolumn{1}{r|}{0.364} &
  0.232 \\ \hline
\end{tabular}%
}
\caption{For each measure, the best value is in bold. PixCorr is a pixel-level correlation score. SSIM is the structural similarity index metric. AlexNet($2$), AlexNet($5$), and AlexNet($7$) are the 2-way comparisons of layers 2, 5, and 7 of AlexNet. CLIP(2-way) is the 2-way comparison of the output layer of the ViT-L/14 CLIP-Vision model. Inception is the 2-way comparison of the last pooling layer of InceptionV3. EffNet-B and SwAV are distance metrics gathered from EfficientNet-B13 and SwAV-ResNet50 models. The brain correlation scores are calculated within the respective regions of the visual cortex. For EffNet-B and SwAV distances, lower is better. Higher is better for all other metrics.}
\label{table:results}
\end{table}

\subsection{Analysis}
\label{analysis}
The improved subjective accuracy performance over the baseline method (our most notable result) appears to be a consequence of our additional brain-aligned optimization inference process. By using a brain encoding model without any pre-training to bias the feature extraction process, our optimization target is trained to be sensitive to the same features as the brain. This builds an intuition for why human raters prefer these images, as presumably, their brains are utilizing similar features to perform their subjective evaluation. Future work could further examine this relationship, exploring whether brain encoding models truly produce embeddings that better approximate human subjective evaluations.

The trajectory of the brain correlation scores over the various stages of the reconstruction pipeline sheds some insights into how our refinement process affects representations in visual cortex. Examining Figure \ref{figure:trajectory}, early visual areas such as V1-V4 have a steeper rate of convergence over the search process, which makes sense given that our brain-optimized inference process refines only low-level features through the updating of the latent $z$ vector. Higher visual areas, by contrast, improve substantially but at a slower rate. The largest improvements in feature metrics over the base decoding method were also found in high-level image metrics, so there appears to also be a clear improvement in the semantic accuracy of the reconstructions. One reason for this could be the specification of a smaller family of images within a CLIP space. Any given CLIP vector can describe a family of images with varying semantic accuracy with respect to the ground truth image, but by selecting images within that space that score well across visual cortex, we narrow the space to contain only images that have a closer semantic relation with the measured brain activity, and by extension the target image.

\begin{figure}[!htb]
\begin{center}
\includegraphics[width=\columnwidth]{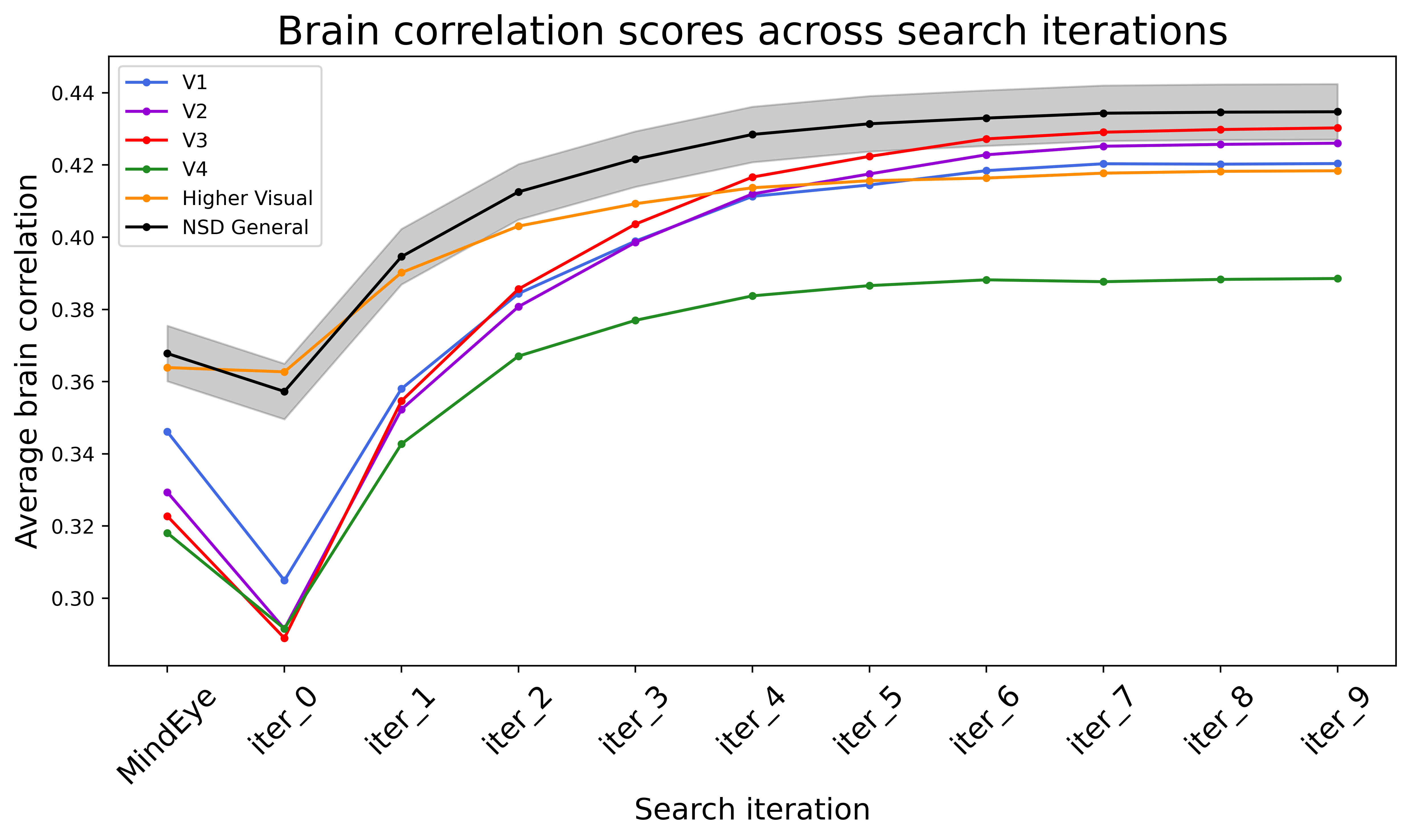}
\end{center}
\caption{Correlation between predicted and actual brain activity (y-axis) for the images at each iteration (x-axis) for different ROIs (curves). For samples that meet the variance target at earlier iterations, the scores from their latest iterations are carried forward and represented in future iterations to avoid a sample bias in the plot.} 
\label{figure:trajectory}
\end{figure}

Figure \ref{figure:alignment} examines the rate at which brain areas reach their target variance threshold, used as our stopping criterion during the BOI algorithm. The percentage of samples that reach this threshold steadily increases across the hierarchy of visual processing areas. Interestingly, the rate at which the samples converge orders brain areas by their level of representational invariance, with V1 converging most slowly, followed by V2, V3, V4, and higher-level visual cortex, which aligns with invariance metrics explored in other works, such as receptive field size \cite{receptivefields}. 

\begin{figure}[!htb]
\begin{center}
\includegraphics[width=\columnwidth]{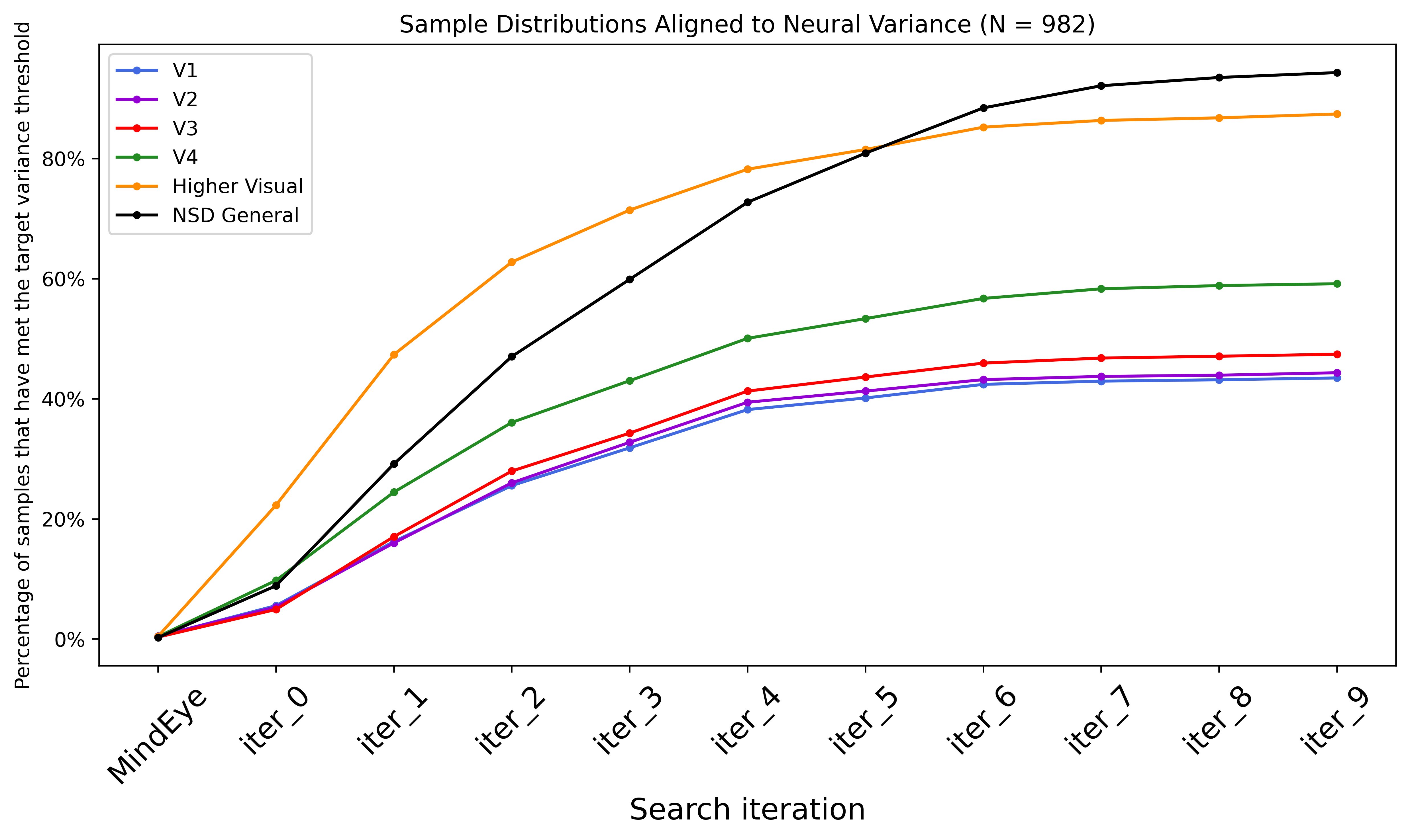}
\end{center}
\caption{Percentage of samples in the test set (n=$982$) for which reconstructions have reached our target variance computed across denoised trial repetitions of the original brain activity (y-axis), for different ROIs (curves) across iterations (x-axis). The target variance for the NSD General mask is used as a stopping criterion during the search, but for this plot, the target variance is computed individually for each brain area.} 
\label{figure:alignment}
\end{figure}

\section{Limitations}

For the technical limitations of our implementation, the most immediate limitation lies in the computational cost of repeatedly sampling images from a diffusion model. This additional inference stage increases the required number of GPU cycles by an order of magnitude relative to the base decoding model. This could be improved with faster image generation models, but will always represent an increase in GPU cycles over direct decoding approaches, as it requires generating many intermediate images in addition to the final image distribution. 

This method also does not have a way to verify the accuracy of the stopping criterion used during the search process. A stopping criterion is necessary to avoid overfitting to the encoding model and our averaged brain activity target, but when representing a distribution, a meaningful measurement of distribution width corresponding to perceptual certainty is also desirable. Our current approach is likely an incomplete solution to many of these objectives, as the autoencoder removes a lot of brain noise that is relevant to quantifying an appropriate estimate of neural uncertainty. Because the encoding model produces noise-free $\beta'$s, the motivation for implementing the autoencoder was to create a mechanism for translating between the encoded beta variance $\sigma(\beta')$ and the measured beta variance $\sigma(\beta)$, which is larger by approximately two orders of magnitude. A more robust and well-motivated stopping criterion would be rich ground for future work, providing a more convincing estimate of the observer's perceptual uncertainty in the represented image distribution.

\section{Conclusion}

We introduce a procedure for optimizing fMRI image reconstructions against a brain correlation metric while simultaneously reducing the breadth of the decoded image distributions to align with brain variance across denoised trial repetitions. We find that this procedure generates high-quality reconstructions and improves performance across a variety of feature spaces without performing any additional optimization in any of them. A key component of our system is our use of brain-optimized encoding models that yield accurate predictions of brain activity in much of visual cortex \cite{St-Yves_heirarchy}, allowing us to leverage the diversity of representations for natural images across visual cortex. The potential of this tool for deeper analysis is also demonstrated in quantifying the invariance in representations across the brain via each brain area’s rate of convergence to our stopping criterion. We find that our method sorts brain areas in agreement with complementary measures of invariance explored in other works. By utilizing encoding models to align reconstructions directly with the measured brain activity, this work aims to expand work in this space to utilize the more descriptive and complete objective of brain correlation, allowing these tools to give new insights into the structure and function of various cortical regions. We believe image reconstruction algorithms have a pivotal role to play in understanding and interpreting brain activity in visual cortex, and reconstructing complex perceptual states.

\section{Acknowledgements}

We would like to thank SHI Labs for providing code and pre-trained models for Versatile Diffusion, and the authors of the MindEye method \cite{scotti2023reconstructing} for providing the base decoding model used in our results. Funding is provided by R01EY023384.

\medskip

\bibliography{main}

\begin{thebibliography}{19}
\providecommand{\natexlab}[1]{#1}

\bibitem[{Allen, St-Yves, and Wu(2022)}]{Allen2021a}
Allen, E.; St-Yves, G.; and Wu, Y. e.~a. 2022.
\newblock A massive 7T {fMRI} dataset to bridge cognitive neuroscience and artificial intelligence.
\newblock In \emph{Nat Neurosci}.

\bibitem[{Gaziv et~al.(2022)Gaziv, Beliy, Granot, Hoogi, Strappini, Golan, and Irani}]{cycleconsistency}
Gaziv, G.; Beliy, R.; Granot, N.; Hoogi, A.; Strappini, F.; Golan, T.; and Irani, M. 2022.
\newblock Self-Supervised Natural Image Reconstruction and Large-Scale Semantic Classification from Brain Activity.
\newblock \emph{NeuroImage}, 254: 119121.

\bibitem[{Gu et~al.(2023)Gu, Jamison, Kuceyeski, and Sabuncu}]{gu2023decoding}
Gu, Z.; Jamison, K.; Kuceyeski, A.; and Sabuncu, M.~R. 2023.
\newblock Decoding natural image stimuli from f{MRI} data with a surface-based convolutional network.
\newblock In \emph{Medical Imaging with Deep Learning}.

\bibitem[{Kay et~al.(2013)Kay, Winawer, Mezer, and Wandell}]{receptivefields}
Kay, K.~N.; Winawer, J.; Mezer, A.; and Wandell, B.~A. 2013.
\newblock Compressive spatial summation in human visual cortex.
\newblock \emph{Journal of Neurophysiology}, 110(2): 481--494.
\newblock PMID: 23615546.

\bibitem[{Kneeland et~al.(2023)Kneeland, Ojeda, St-Yves, and Naselaris}]{kneeland2023reconstructing}
Kneeland, R.; Ojeda, J.; St-Yves, G.; and Naselaris, T. 2023.
\newblock Reconstructing seen images from human brain activity via guided stochastic search.
\newblock Conference on Cognitive Computational Neuroscience.

\bibitem[{Lin, Sprague, and Singh(2022)}]{lin2022mind}
Lin, S.; Sprague, T.~C.; and Singh, A. 2022.
\newblock Mind Reader: Reconstructing complex images from brain activities.
\newblock In Oh, A.~H.; Agarwal, A.; Belgrave, D.; and Cho, K., eds., \emph{Advances in Neural Information Processing Systems}.

\bibitem[{Lin et~al.(2014)Lin, Maire, Belongie, Hays, Perona, Ramanan, Doll{\'a}r, and Zitnick}]{microsoftcoco}
Lin, T.-Y.; Maire, M.; Belongie, S.; Hays, J.; Perona, P.; Ramanan, D.; Doll{\'a}r, P.; and Zitnick, C.~L. 2014.
\newblock Microsoft COCO: Common Objects in Context.
\newblock In Fleet, D.; Pajdla, T.; Schiele, B.; and Tuytelaars, T., eds., \emph{Computer Vision -- ECCV 2014}, 740--755. Cham: Springer International Publishing.
\newblock ISBN 978-3-319-10602-1.

\bibitem[{Lu et~al.(2023)Lu, Du, Wang, and He}]{lu2023minddiffuser}
Lu, Y.; Du, C.; Wang, D.; and He, H. 2023.
\newblock MindDiffuser: Controlled Image Reconstruction from Human Brain Activity with Semantic and Structural Diffusion.
\newblock arXiv:2303.14139.

\bibitem[{Naselaris et~al.(2011)Naselaris, Kay, Nishimoto, and Gallant}]{naselaris2011}
Naselaris, T.; Kay, K.~N.; Nishimoto, S.; and Gallant, J.~L. 2011.
\newblock Encoding and decoding in {fMRI}.
\newblock \emph{NeuroImage}, 56(2).

\bibitem[{Ozcelik and VanRullen(2023)}]{ozcelik2023braindiffuser}
Ozcelik, F.; and VanRullen, R. 2023.
\newblock Brain-Diffuser: Natural scene reconstruction from {fMRI} signals using generative latent diffusion.
\newblock arXiv:2303.05334.

\bibitem[{Radford et~al.(2021)Radford, Kim, Hallacy, Ramesh, Goh, Agarwal, Sastry, Askell, Mishkin, Clark, Krueger, and Sutskever}]{radford2021learning}
Radford, A.; Kim, J.~W.; Hallacy, C.; Ramesh, A.; Goh, G.; Agarwal, S.; Sastry, G.; Askell, A.; Mishkin, P.; Clark, J.; Krueger, G.; and Sutskever, I. 2021.
\newblock Learning Transferable Visual Models From Natural Language Supervision.
\newblock In Meila, M.; and Zhang, T., eds., \emph{Proceedings of the 38th International Conference on Machine Learning}, volume 139, 8748--8763. PMLR.

\bibitem[{Rombach et~al.(2021)Rombach, Blattmann, Lorenz, Esser, and Ommer}]{stablediffusion}
Rombach, R.; Blattmann, A.; Lorenz, D.; Esser, P.; and Ommer, B. 2021.
\newblock High-Resolution Image Synthesis with Latent Diffusion Models.
\newblock \emph{CoRR}, abs/2112.10752.

\bibitem[{Scotti et~al.(2023)Scotti, Banerjee, Goode, Shabalin, Nguyen, Cohen, Dempster, Verlinde, Yundler, Weisberg, Norman, and Abraham}]{scotti2023reconstructing}
Scotti, P.~S.; Banerjee, A.; Goode, J.; Shabalin, S.; Nguyen, A.; Cohen, E.; Dempster, A.~J.; Verlinde, N.; Yundler, E.; Weisberg, D.; Norman, K.~A.; and Abraham, T.~M. 2023.
\newblock Reconstructing the Mind's Eye: fMRI-to-Image with Contrastive Learning and Diffusion Priors.
\newblock In \emph{37th Conference on Neural Information Processing Systems}.

\bibitem[{St-Yves et~al.(2022)St-Yves, Allen, Wu, Kay, and Naselaris}]{St-Yves_heirarchy}
St-Yves, G.; Allen, E.~J.; Wu, Y.; Kay, K.; and Naselaris, T. 2022.
\newblock Brain-optimized neural networks learn non-hierarchical models of representation in human visual cortex.
\newblock \emph{bioRxiv}.

\bibitem[{St-Yves and Naselaris(2019)}]{St-Yves_gan}
St-Yves, G.; and Naselaris, T. 2019.
\newblock Generative Adversarial Networks Conditioned on Brain Activity Reconstruct Seen Images.
\newblock In \emph{2018 IEEE International Conference on Systems, Man, and Cybernetics (SMC)}, 1054--1061. Institute of Electrical and Electronics Engineers Inc.

\bibitem[{Takagi and Nishimoto(2023{\natexlab{a}})}]{Takagi2022.11.18.517004}
Takagi, Y.; and Nishimoto, S. 2023{\natexlab{a}}.
\newblock High-resolution image reconstruction with latent diffusion models from human brain activity.
\newblock In \emph{Proceedings of the IEEE/CVF Conference on Computer Vision and Pattern Recognition}, 14453--14463.

\bibitem[{Takagi and Nishimoto(2023{\natexlab{b}})}]{takagi2023improving}
Takagi, Y.; and Nishimoto, S. 2023{\natexlab{b}}.
\newblock Improving visual image reconstruction from human brain activity using latent diffusion models via multiple decoded inputs.
\newblock arXiv:2306.11536.

\bibitem[{Van~Bergen et~al.(2015)Van~Bergen, Ma, Pratte, and Jehee}]{sensoryuncertainty}
Van~Bergen, R.; Ma, W.; Pratte, M.; and Jehee, J. 2015.
\newblock Sensory uncertainty decoded from visual cortex predicts behavior.
\newblock \emph{Nature neuroscience}, 18.

\bibitem[{Xu et~al.(2023)Xu, Wang, Zhang, Wang, and Shi}]{Xu_2023_ICCV}
Xu, X.; Wang, Z.; Zhang, G.; Wang, K.; and Shi, H. 2023.
\newblock Versatile Diffusion: Text, Images and Variations All in One Diffusion Model.
\newblock In \emph{Proceedings of the IEEE/CVF International Conference on Computer Vision (ICCV)}, 7754--7765.

\end{thebibliography}

\end{document}